\documentclass[11pt,a4paper]{article}
\usepackage{amsmath, amssymb, verbatim, graphicx, color}
\usepackage{fullpage}

\begin{document}

\title{Random ``choices'' and the locality loophole}
\author{Stefano Pironio\\[0.5em]
Laboratoire d'Information Quantique, Universit\'e Libre de Bruxelles (ULB), Belgium}

\date{September 30, 2015}

\maketitle

The results of any Bell experiment can in principle be explained by a hidden-variable model if information about the measurement settings on one side can travel to and influence the measurement on the other side. To exclude this possibility, or at least make it incompatible with special relativity, the entangled particles should be sufficiently distant from each other and the measurement settings varied sufficiently rapidly, so that such information should necessarily travel at a speed greater than $c$ to have any influence on the measurement results. Failure to satisfy this condition is known as the ``locality loophole".

In addition to this space-like separation condition, attention should also be payed to the mechanism that is used to vary the measurement settings on each side. Indeed if this process is not truly random, information about its outcome is in principle already available somewhere in the past, maybe even before the start of the experiment, and could therefore have had plenty of time to influence the behavior of the entangled particles.

The opinion is held among some physicists that the measurement settings \emph{should} therefore be chosen using a random process of quantum origin. This has first been implemented this way in the famous Weihs et al experiment \cite{weihs_violation_1998}, and afterwards in all recent Bell experiments \cite{scheidl_violation_2010,erven_experimental_2014,hensen_experimental_2015} addressing the locality loophole.

But isn't it a little bit circular to rely on quantum randomness to test local hidden-variable models according to which quantum effects might be completely deterministic? And if from the perspective of such local hidden-variable models, the randomness of a quantum measurement is no different than the pseudo-randomness of a classical coin toss, why not use such classical processes to select the measurement setting? If it is conceptually equivalent, wouldn't it be at least simpler from the implementation point of view?

I point out here that, indeed, quantum random number generators (QRNGs) are unnecessary for testing the violation of Bell inequalities. A Bell experiment based on certain pseudo-random -- but otherwise completely deterministic --  mechanisms for selecting the measurement settings, such as taking a hash function of the latest $10^6$ tweets with the hashtag $\#$quantum, would be as convincing, if no more, than a Bell experiment based on QRNGs.

Before proceeding, let me stress that I'm considering here Bell experiments as fundamental tests of Nature, that is, from the perspective of ruling out local hidden-variable models (or local realism or local causality depending on one's terminology). The arguments presented below do not necessarily directly apply to other uses of Bell inequalities, as, e.g., in device-independent quantum information where one usually assumes the validity of quantum theory as a premise. I should say, though, that I also think that the use of QRNGs is not fundamentally required for device-independent applications, but not exactly for the same reasons as the ones discussed below.

\section{Locality and randomness in an ideal situation}\label{sec:ideal}
Before discussing in detail the conditions needed to realize a proper Bell experiment, it may be useful to consider a simpler (but also more artificial) situation. Suppose that Alice and Bob are two quantum-deniers who claim to have found a local hidden-variable model reproducing the predictions of quantum theory. We challenge them to produce a Bell inequality violation using a simulation of their local hidden-variable model. We thus place Alice and Bob in two separate rooms, equipped with classical computers for running their simulation.  We give each of them an input -- corresponding to a measurement setting --, they run their simulation, and after some time they each give us back an output -- corresponding to a measurement outcome. We repeat this process $N$ times and at the end we analyze the sequence of inputs and outputs to verify whether a violation, say, of the CHSH inequality, is obtained.

No matter how ingenuous is Alice and Bob's model and no matter how powerful are their computers, we can make the probability that they pass the challenge as small as we wish by making the number of runs $N$ sufficiently large. This is true provided that two conditions hold:
\begin{itemize}
\item A \emph{no-communication or locality} condition: in each run, Alice and Bob cannot communicate to each other their respective inputs before generating their outputs. (They can, though, communicate freely between runs if they wish).
\item A \emph{randomness} condition: the inputs that we give to them must be random. By this, I mean that Alice and Bob cannot predict the inputs that they will receive and thus adapt their strategy to the future values of the inputs.
\end{itemize}
These are the only restrictions that we need to put on Alice and Bob. If they are satisfied, there is no way for them to win the challenge, or at most only with vanishingly small probability.

It is not too difficult to imagine ways to set-up the challenge so that these two conditions are satisfied. To guarantee the no-communication condition, we can for instance arrange things such that Alice's and Bob's rooms are completely isolated from each other and the outside world, and such that the only way that they can communicate with us is through a small hatch in the door of their rooms through which a piece of paper can be passed with the inputs or the outputs written down. As regards the randomness condition, we could choose the inputs using sufficiently complex physical processes or cryptographically-secure pseudo-random generators. It is highly unlikely that Alice and Bob, whose physical and mathematical knowledge seems quite limited and who are confined in their isolated rooms, will be able to predict our future input choices based only on the value of past ones -- at least not with sufficient accuracy to win the challenge. But if we are afraid that that this might happen, we could as well simply use QRNGs. Indeed, there is no single physicist today that can predict the output of a well-conceived QRNG and that's most likely also true for Alice and Bob. 

In this idealized setting, a proper Bell test is thus based on two well-defined conditions. These two conditions are logically independent: there may be situations in which the no-communication condition is satisfied but not the randomness one (say, because we systematically alternate the input values from one run to the other) and situations in which the randomness condition is satisfied but not the no-communication one (say, because we forgot to confiscate Alice's and Bob's cell phones). 

\section{Locality and randomness in real Bell tests}\label{sec:real}
Roughly speaking, one could identify in the context of real Bell experiments the violation of the no-communication condition with what people call the locality loophole and the violation of the randomness condition with what people call the measurement-independence or freedom-of-choice loophole. I say ``roughly speaking'' because in a real Bell experiment things are not as neat as in the above artificial situation. In a real Bell experiment, one where we are actually testing if Nature can be described by a local hidden-variable model and where we do not assume the validity of quantum theory, we are not merely dealing with classical physicists running a simulation in a well-controlled environment, but with real entangled particles which can, from the perspective of a local hidden-variable model, behave in ways that are completely beyond our control.

First of all, we cannot be sure that what looks random to us is indeed random from the point of view of the two entangled quantum particles. Even the output of a QRNG may not be random since it may also be described by hidden-variables and these hidden-variables could be correlated to the ones describing the entangled particles.

Second, how do we know that the two particles are not communicating? We could place the two measuring apparatuses in distant, shielded boxes. But the two particles might nevertheless be communicating through some (subluminal) hidden signals of which we are presently unaware and which we cannot block with current technology. One natural solution to this problem is to set-up the experiment in such a way that the event corresponding to ``the input being given" on one side is space-like separated from the event corresponding to ``the output being obtained" on the other side. But then this leads to two new problems.

First, how do we know when ``an output is obtained"? Quantum theory does not tell us unambiguously when the collapse of the wavefunction happens. Is it when the observed photon triggers an electron avalanche? When a computer registers the output of the photo-detector? When the measurement result induces a massive object to move? When the measurement outcome is read by a human observer? This ambiguity in the definition of the event associated with the registration of a measurement outcome and the associated potential failure of the space-like separation condition is known as the collapse locality loophole \cite{kent_causal_2005,salart_spacelike_2008}. In the following, we will ignore this loophole and assume for simplicity that we can define the time at which an output is obtained.

We are still faced with another problem though, which is when can we consider that ``the input is given"? The particles are not isolated in some room with a little hatch in the door that we can open to give the input. When should we then consider that potential information about the measurement setting on one side can start propagating to the other side? When the quantum particle reaches the measuring apparatuses? When a knob is turned on the measuring apparatus? When our random number generator ``chooses" a measurement setting? But then when is this random choice made? And if the generator is not actually random, how can we even define when a choice is made?

We thus see that in a real Bell experiment determining whether the no-communication and randomness conditions are satisfied becomes much more ambiguous. Moreover, while the no-communication and randomness conditions were logically independent, this does not seem to be longer the case. More problematically, we can never be sure that these two conditions are satisfied. No matter how sophisticated is the experiment, one can always concoct a local hidden-variable model that explains it. Logically, it is not possible to ``close'' the locality loophole and the measurement-independence loophole (in the same way that the detection loophole can be closed). 

How should we then decide what counts as a satisfactory Bell experiment? Let us approach this question as if we were ignorant of the way that it has been framed in recent experiments.

\section{Ruling out a family of physically plausible models}\label{sec:plausible}
We cannot rule out every local hidden-variable model that is \emph{logically possible}, but we should at least aim to rule out every model that is \emph{physically plausible}. This is the best that we can do. As professional physicists, this is also the only thing that should matter to us.

To understand how to look at things from this perspective, let us consider what would be  required to rule out a simple, but natural, family of local hidden-variable models. From there, we will then move to more complex and general models. To be concrete, let us also discuss this issue on the paradigm of an optical Bell experiment. We thus assume that a type-II parametric down-conversion source produces pairs of photons entangled in polarization.  Each photon is then directed to a distant measurement station, where it goes through an orientable wave-plate (used for selecting the measurement basis) followed by a polarization beam-splitter, and is finally detected in one of the beam-splitter exit ports. 

As a starting point, consider the following very simple family of local hidden-variable models for this experiment. We assume that the measurement outcomes are determined by some hidden-variables $\lambda$ with distribution $q(\lambda)$. We think of $\lambda$ and $q(\lambda)$ as parameters characterizing the source of entangled photons. They thus depend, e.g, on the type of laser and non-linear crystal that are used to produce the entangled photons. Given a specific value for $\lambda$, the local hidden-variable model then assigns a probability for each possible outcome when a measurement is made on photon $A$ and a probability for each possible outcome when a measurement is made on photon $B$. We assume that the outcome probabilities for each photon depend, in addition to $\lambda$, on the characteristics of the components (wave-plate, polarizing beam-splitter, photo-detectors, etc) from which the measurement apparatus is made and in particular, of course, on the wave-plate orientation at the time that the photon goes through it. 

In principle, distant parts of the experiment could (subluminally) influence each other arbitrarily, and thus $\lambda$ could also depend, e.g., on the orientation of the wave-plate at a remote measurement station at some earlier time. The particular mechanisms that we use to select the measurement settings could also influence the measurement outcomes. For instance, we could rotate the wave-plates manually based on the outcome of a coin toss or use motorized rotation mounts controlled by the output of pseudo-random number generator running on a remote computer. Even if the wave-plates are oriented exactly in the same direction, the local hidden-variable model could nevertheless predict different outcome probabilities in the two cases. Finally, the outcomes probabilities could not only depend on the physical configuration of the measurement apparatuses at the moment at which the photons go through them, but also on the history at each measurement station. For instance, in a given experimental run, and all other factors being equal, different predictions could be obtained if we had being measuring the photons hundred times in a row in the same direction or if we had being randomly switching between orientations.

But let us assume, at least in a first step, that these possibilities do not arise and that the photons behave in the simple fashion outlined above. Note that this actually represents a very natural family of local hidden-variable models. For instance, when we describe a Bell experiment according to quantum theory, we usually adopt the same point of view. In quantum theory, we assign a state $|\psi\rangle$ to the entangled particles. To determine $|\psi\rangle$ we need to take into account the properties of the source, such as the type of laser and crystal that are used, but we usually do not need to take into account the experimental configuration at the distant measurement stations or what kind of measurements we are going to apply later. Similarly, given $|\psi\rangle$ the only property of the measurement apparatus that matters to determine the probability to get a certain outcome is the type of measurement that we apply at the moment that we apply it. But details such as the type of mechanism that we use to select the measurement or what kinds of measurements were made before are irrelevant. In principle, it is true that, say, whether we rotate a wave-plate manually or whether we use a motorized rotator controlled by a computer will have a different effect on the outcome probabilities, even if in both cases the wave-plate is oriented in the exact same direction. This is because in addition to the wave-plate, the polarizing beam-splitter, and the photo-detectors themselves, the atomic and electronic configuration of other nearby objects also have an influence on the photon. We can even in principle compute the effects that such tiny experimental details have on the outcome probabilities. But they are so negligible that they will have no observational consequences and thus we usually ignore them when we model a Bell experiment in quantum theory. 

It is thus very natural to make, at least in a first attempt, the same assumptions for our local hidden-variable models. Indeed, while they are various motivations for introducing such models, they all revolve around the idea that some aspects of the mathematical formalism of quantum theory in term of the wave-function $|\psi\rangle$ and the Born rule are not adequate. We thus seek an alternative formalism based on some other parameters $\lambda$. These parameters may not yet be known or accessible, but this does not necessarily imply that they should describe arbitrary hidden influences between every little aspect of the experimental set-up, or at least not in a way that has some observational significance. If we were to take such an approach each time we have to build a model for a physical experiment or update a theory, we would never go very far. 

A simple local hidden-variable model of the above sort can readily be concocted if entanglement does not survive when the photons separate and if instead the joint state factorizes in a random product state, a hypothesis introduced by Schr\"odinger \cite{schrodinger_discussion_1935} and Furry \cite{furry_note_1936}. Such a possibility was apparently taken seriously by many physicists until the seventies \cite{clauser_bells_1978}. 

Since the predictions of these simple models only depend on the local wave-plate orientation at the moment at which the measurement is made (but not on the mechanism used to determine this orientation or the orientations in previous runs) to rule them out experimentally it is sufficient to use static orientations. That is, the different orientations necessary to test a Bell inequality can be kept fixed and varied only after a whole bunch of measurement runs have been registered. Experiments of this kind have been done in the seventies, and the experimental results follow the quantum predictions and refute our simple local hidden-variable models.

What is the next step, then? Physicists have come-up with hidden-variable models which can nevertheless explain the results of such experiments if some communication is going on between the two entangled particles or the measurement devices. Examples are Bohm's theory \cite{bohm_suggested_1952} or collapse models were the measurement on one half of an entangled pair induces a (delayed) update of the local state of the second half \cite{eberhard_realistic_1989}. 

So let us now add this new ingredient and let us assume that the outcome probabilities of particle $A$ may also depend, in addition to $\lambda$ and the local wave-plate orientation, on some message $m$ (carried by a wave or a hidden particle) coming from the distant measurement station $B$. Similarly, the output at $B$ may depend on a message coming from $A$.  But adding this new ingredient to the model does not forces us to give up our natural assumptions about what is relevant and irrelevant to describe the physics of the experiment. So keeping in line with the previous discussion, we assume that the message $m$ sent from side $B$ at time $t$ only depends on the orientation of the wave-plate on side $B$ at that time. A model of this sort is already perfectly sufficient to reproduce the predictions of quantum theory.

Note that in principle, the message $m$ sent from side $B$ could carry arbitrary information about what is happening on that side, for instance $m$ could already convey information about a future wave-plate orientation based on the present initial conditions of the coin that will later be tossed to determine that orientation. This piece of information could then be exploited by particle $A$ to determine its measurement output.  But we assume for the moment that this is not the case. Again, think about how a physicist would model a real Bell experiment using Bohm's theory or some explicit mechanism for the collapse of the wave-function.

Models of the above sort cannot be ruled out in a standard Bell experiment if the hidden communication travels at an arbitrary speed.  But if we assume that this communication travels at a speed lower or equal to $c$, we recover a notion of locality and we can at least hope to rule out that possibility. Note that if communication propagates at a finite speed, changing the orientation of the wave-plate on one side can have an influence on the other side only after a certain time lapse. But even if the information sent from side $B$ arrives on side $A$ with a delay, if the wave-plate orientation on side $B$ has not been changed in the meantime, then the information received at $A$ correctly describes the present situation at $B$. Thus such hidden-variable model involving subluminal communication can explain the results of Bell experiments performed with static measurements and to rule them out we are led to consider experiments where, in the worlds of Bell, ``the settings are changed during the flight of the particles'' \cite{bell1}. In the context of the models that we have been considering so far, closing the locality loophole thus requires a space-like separation between the moment at which the wave-plate is rotated on one side and the moment at which an outcome is registered on the other side. If this is the case, any communication explaining the measurement results must propagate faster than light. 

How should we change the orientation of the wave-plates? We should be a little bit careful. For instance in the famous experiment by Aspect et al \cite{aspect_experimental_1982}, periodical switching between the two possible measurement settings occurred every $10 \text{ns}$, which was shorter than the time of $40 \text{ns}$ for a light signal to travel between the two wings of the experiment. As pointed out in \cite{zeilinger_testing_1986}, however, the time for a light signal to go from $A$ to $B$ was, by accident, a multiple of the switching period. It follows, that a message $m$ propagating at the speed of light and leaving the measurement station $B$ would arrive at $A$ when the measurement setting at $B$ had been returned to its value at the time of emission. Thus the messages received at $A$ at a given time would correctly describe the situation at $B$ at the same time, in which case an underlying model may obviously reproduce the quantum predictions. To avoid any issue of this sort, we could nowadays simply switch pseudo-randomly the settings using some complicated function generated by a computer. This was not possible at the time of the experiment by Aspect et al, but should be doable with current technology. 

\section{Ruling out general models, modulo improbable conspiracies}\label{sec:conclusive}

The models that we have been considered so far may seem too simple or limited and one may want to rule out more general family of local hidden-variable models, according to which aspects of the experimental set-up that we have neglected actually play a significant role.  But they illustrate two important points. 

First, according to any given local hidden-variable model, there are aspects of the physical world that are relevant to describe the experiment -- e.g. the laser and non-linear crystal used to generate the entangled pair and the orientation of the wave-plates in the two measurement stations -- and they are aspects of the physical world that are largely irrelevant (i.e. do not have significant observable effects), once we have specified the relevant ones. For instance in the models considered above, the type of mechanism used to rotate the wave-plates and the pseudo-random processed used to choose their orientations had obviously an effect on the orientations of the wave-plates themselves, but apart from that were useless pieces of information that were not exploited by the quantum systems to determine their outputs.  Different models will obviously differ in what counts as relevant or irrelevant, but any Bell test necessarily rules out local hidden-variable models according to such a distinction.

Second, in contrast to the challenge against Alice and Bob that we have been discussing in Section~\ref{sec:ideal}, in which the measurement settings had to be changed from one run to the other because of the randomness condition -- because otherwise Alice and Bob could have easily adapted their strategies to our fixed measurement choices -- this is not the reason why we have been led to introduce Bell experiment with varying measurement settings. Indeed, in the simple models that we have just discussed, the local behavior of the particles only depends on the orientation of the wave-plates at a given time, but is independent of their previous orientations: for a given value of the hidden-variable $\lambda$, the local wave-plate orientation $w$, and the message $m$ received from the other side, the particles will behave in the same manner whether the wave-plates have been kept fixed for one hundred runs or whether they have been randomly changed. We have instead been led to consider Bell experiments with varying settings because of the locality condition, because this was necessary to rule out the possibility that the observed data could be explained by the subluminal communication $m$ between the two measurement devices.

Specifically, closing the locality loophole from the perspective that we have been considering requires a space-like separation between what counts as a ``change in the measurement setting'' on one side and the registration of an outcome on the other side. Similarly to the challenge against Alice and Bob, there is thus a sense in which the locality condition can be addressed independently of the randomness one. What counts as a change in the measurement setting -- informally the time at which ``an input is given to the quantum system'' -- depends on the parts of the experimental set-up that may have a relevant effect on the behavior of the quantum systems (according to some specific model). In the example discussed above, ``an input was given to the quantum system'' when we changed the orientation of the wave-plate.  More general models could be sensitive to more subtle aspects of the experimental set-up, and it could be that an ``input is given'' for instance already at the time at which a signal specifying the measurement orientation is sent by the electronics that controls the measurement devices. 

Of course there is a lot of arbitrariness involved in the distinction between what counts as relevant, what counts as irrelevant, and in the associated definition of the events that are used to define the space-like separation necessary to address the locality loophole. But
\begin{enumerate}
\item We necessarily need to introduce such a distinction, because as we have said earlier, if everything can have some relevant effect on our experiment, then there is always a local hidden-variable model that explains it, no matter how sophisticated is the experiment.
\item For all practical purposes, the fact that a Bell experiment only rules out local hidden-variable model relative to such a distinction is perfectly acceptable if it would be ludicrous to question it to explain the results of the experiment.
\end{enumerate}
For instance consider an experimental set-up, which apart from the way the measurement settings are chosen, is identical to the one of Hensen et al \cite{hensen_experimental_2015} based on entangled NV center electron spins in diamond. Suppose that the entire sequence of $N$ pairs of future measurement settings is determined just before the experiment by taking a complex hash function of the $10^6$ most recent tweets with the hashtag $\#$quantum. The sequence is divided in two subsequences, consisting each of $N$ blocks of 10 digits, and each subsequence is  stored in a ``randomness box'', one for each system. At each measurement run $k$, the parity of the 10 digits of the $k^\text{th}$ block in the sequence on each side is computed and is used to set the state of a fast microwave switch which selects the measurement basis. Suppose further that the beginning of the space-like interval required to close the locality loophole is set 200 ns before the parity computation.  Then I think that such an experiment -- let us refer from now on to any experiment of this sort as a Twitter-based experiment\footnote{Of course the use of Twitter is purely illustrative and plays no special role. The content of the random boxes could be generated in any other sufficiently complex way unlikely to be correlated to the quantum particles.} -- would represent a conclusive Bell test convincingly addressing the locality and measurement-independence loopholes.

Strictly speaking such an experiment only rules out a large class of local-hidden variable models, those for which the behavior of the observed systems is not in any relevant way influenced by or correlated to the content of the randomness boxes. It is true that it is logically possible that the same multitude of factors that are responsible for the content of the latest $10^6$ tweets with the hashtag $\#$ quantum also have a key influence on the behavior of our entangled particles, and that this influence is responsible for the observed violation, even after taking into account that the tweets have been processed by a hash function. Or, since before each measurement run, the decision of what measurement basis is going to be used on particle $A$ is already written well in advance in 10 bits of the memory of the ``randomness box'', it could be that information about these 10 bits -- and only about these ones out of the thousands of other random bits stored in the box -- propagates at subluminal speed to the other side of the experiment and his exploited by particle $B$ to determine its measurement output. But, quoting Bell,
\begin{quote} This way of arranging quantum mechanical correlations would be even more mind boggling than one in which causal chains go faster than light.  \cite{bell2}
\end{quote}
\begin{quote}
A theory may 
appear in which such conspiracies inevitably occur, and these conspiracies 
may then seem more digestible than the non-localities of other theories. 
When that theory is announced I will not refuse to listen, either on 
methodological or other grounds. But I will not myself try to make such 
a theory. \cite{bell_exchange_1985}
\end{quote}
 
\section{Superlocality: an extreme interpretation of locality}\label{sec:superloc}
Nothing that I have said above is particularly original\footnote{See for instance \cite{bell_exchange_1985}, where Bell, Clauser, Horne, and Shimony say essentially the same thing, though in a much more clear, concise, and sharp way (as I should have realized before written the present notes).}  nor, I think, controversial. Yet the issue of how to address the locality loophole and how to choose the measurement settings is not considered in those terms in modern experiments. I have explicitly been told by some physicists, who are perfectly reasonable fellows for all other purposes, that an experiment similar to the Twitter-based one \emph{would not} constitute a conclusive Bell experiment and would leave the locality loophole open. 

The way that they frame this issue usually follows the papers \cite{weihs1} and \cite{weihs_violation_1998} of Weihs et al. In \cite{weihs_violation_1998}, the following definition of the locality condition is introduced:
\begin{quote}
The assumption of locality in the derivation of Bell's theorem requires that the individual measurement processes of the two observers are spacelike separated. We define an individual measurement to last from the first point in time which can influence the choice of the analyzer setting until the final registration of the photon.
\end{quote}
According to this definition, what matters to close the locality loophole is the first point in time at which some influence can affect the \emph{choice} of measurement setting. What counts is not the moment at which the measurement setting is \emph{physically adjusted} to some particular value (e.g. by rotating a wave-plate), but the first moment in time at which this value is chosen and recorded somewhere in the physical world. In particular \cite{weihs1} 	,
\begin{quote}
Any reading of the random numbers from a pre-existing list or even calculation of the random numbers using a pseudo-random generator could not enforce locality because in both of these cases the information about which numbers will be generated and possibly used for analysis is already present at the beginning of the experiment.
\end{quote}
This therefore excludes experiments such as the one proposed above based on Twitter-RNGs. One needs true random numbers to close the locality loophole, freshly generated at each single measurement run, and since the only source of true random numbers that we know of is of quantum origin, QNRGs should be used to select the measurement settings. This is the rationale behind the use of QRNGs in all modern Bell experiments that have addressed the locality loophole. 

In the following, let me start by pointing out that from a physical point of view, the Weihs et al definition is not a satisfactory way to look at the locality loophole. Remark first that it is a very different perspective than the one I advocated above, where to define the space-like separation between the two wings of the experiment, what actually matters is not the moment at which the choice to apply a given setting or another is made, but the moment at which we physically change, e.g., the orientation of a wave-plate, or more generally some part of the experimental set-up which can plausibly have some significant influence on the behavior of the observed systems.

According to Weihs et al instead, everything that is not explicitly ruled out by special relativity, every little piece of information that could reach at subluminal speed the quantum particles, is assumed to open the locality loophole, independently of whether there is any good physical reason to think that this information has indeed any relevant effect on the systems that we are observing. This is clearly a very extreme and conspiratorial interpretation of the locality condition. In analogy with superdeterminism, I propose to call it \emph{superlocality}.

Again, let me stress that this does not corresponds at all to the way that professional physicists usually think when they try to build models of the natural world. If every time that we want to understand what happens to some physical system, what kind of physical laws determine its behavior, we had to consider the entirety of the events in its past-light cone we wouldn't go anywhere. Yet, this has somehow become the default view among a certain circle of physicists as far as the interpretation of Bell experiments is concerned. 

It is easy to understand why. At first sight, superlocality seems a much better criterion to address the issue of locality than the one we used in Section~\ref{sec:conclusive}. In our earlier discussion, there was some arbitrariness involved in the definition of the space-like interval because there was some arbitrariness in what it means to change a measurement setting: Is it when the wave-plate is rotated? When a mechanical knob is turned on the measurement device? When a distant computer sends its instructions? Now, this arbitrariness is removed as the only thing that matters is the moment at which the choice of the measurement setting is made.

It is true that this would be a better way to define the space-like interval if we had a true source of randomness, if we had a magic box guaranteed to produce outputs at random, even with respect to local hidden-variable models. The problem is that we do not have such a magic box. Whatever appears random to us, even the output of a QRNG, could in principle be completely determined a long time ago by some hidden-variables and thus if we interpret superlocality strictly we can never satisfy it, almost by definition.

Of course, people realized this. Thus when using the output of a QRNG to define the space-like interval in a Bell experiment, an additional assumption is made that the output of the QRNG is truly random, as explicitly stated in \cite{scheidl_violation_2010} and \cite{hensen_experimental_2015}, or more generally that whatever determines the output of the QRNG is irrelevant for the purpose of explaining the behavior of the observed system. But then, isn't this additional assumption equivalent to a statement about what should be considered relevant and irrelevant and thus when looked from this perspective isn't superlocality not much different in practice from the way we have been looking at the locality condition in the previous Section?

The answer is no because superlocality never confronts directly the central question which is whether the assumptions that have to be necessarily introduced to rule out local hidden-variable models are physically plausible. Or better said, whether questioning these assumptions to maintain a local hidden-variable view would be ludicrous. Superlocality replaces this central question with a ready-made ideological answer, which is that the only things that matters is the moment at which the measurement choice is determined. But this answer is useless because the moment at which this choice is made is often either ill-defined or not known.

Let me give an example to illustrate that. Many people consider that the ultimate Bell experiment (or at least a very convincing Bell experiment) would be one where space-like separated human observers directly change the measurement settings. Such an experiment would look like this. Alice is on earth, Bob is sent very far away on a space ship, say at 10 light-seconds. In each run of the experiment, a light is switched on in Alice's lab and one in Bob's lab for an interval of a few seconds. During that interval, Alice and Bob each have to make a random choice -- 0 or 1 -- corresponding to two possible measurement settings – and turn accordingly the knob on their measurement apparatus. If they do this while the light is on, it is guaranteed that no information about their measurement choice can travel at light speed to the other side before an outcome is registered.

I think that almost everyone would agree that this indeed qualifies as a very convincing experiment. But now consider the following situation. Two seconds before her light is on, Alice chooses a random setting, 0 or 1, quickly writes it on a piece of paper and then uses this choice to determine how she turns her knob when the light is on. The only difference between this new situation and the previous one as regards the arrangement of matter in the physical world are a few molecules of ink on a sheet a paper. Do we really think that this would make a difference to the interpretation of the experiment? I think that one would have to be insane to argue that yes, there is a physically plausible and viable path for local hidden-variable models in one case but not in the other.
Yet, according to the superlocality condition, the first experiment is OK, but the second is not because Alice's measurement choice is now written on a piece of paper and is no longer space-like separated from the other side. Remember that according to superlocality ``any reading from a pre-existing list'' is \emph{by definition} rejected.

We can even invent more ludicrous situations. Suppose that ten seconds before the light is on, Alice writes ten numbers on her sheet of paper, but always use the third number to decide how she turns the knob. Again this fails the superlocality condition as the choice of measurement is already written down somewhere in the physical world outside the required space-like interval and could therefore influence the entangled particles. But how come only the physical configuration of the ink molecules forming the third number, and not the other ones, have such an influence? That sounds completely crazy. Actually, Alice does not even need to write down her number. She can simply make her choice a few ms before the light is on and keep it in her head. Again, her choice is recorded in a few synaptic connections in her brain outside the required space-like interval and thus the superlocality condition is not satisfied.

As you see, if we interpret strictly the condition of superlocality, we cannot close the locality loophole using human observers; we cannot make the ultimate Bell experiment.  We would always worry that the observers consciously or unconsciously make their choice before they are allowed to do so.  On the other hand, if we look instead at things from the perspective of what is physically plausible or implausible, what can have a relevant or irrelevant effect on the outcome of the experiment, it does not really matter that the human observers make their choice a few ms before the event defining the space-like interval. It seems very reasonable to assume that the detection outcomes are not influenced by the few neurons\footnote{There are interpretations of quantum theory that explicitly assume that our minds have something to do with the collapse of the wave-function. But these theories are already out of the scope of  local hidden-variable models, whose aim is after all to come-up with some reasonable and ``classical" explanations for quantum effects. Furthermore those interpretations do not generally claim to restore locality to quantum theory.} (out of the 100 billions neurons of the human brain) that have registered the measurement choice or by a few molecules of ink on a piece of paper\footnote{Pushing things a little bit further, one could even say that with present technology we can already do a Bell experiment where the choice of settings is determined by two free-willed experimentalists. We could simply ask Alice and Bob to generate in advance a list of random choices and store them in two ``random boxes" determining the measurement settings in a way similar than in the Twitter-based experiment.}. But if we find it reasonable to look at things from this perspective in the context of Bell experiments involving human observers, there is no reason to proceed differently in more practical experiments.

In summary, the superlocality criterion is not a satisfactory nor useful way to look at the locality loophole. Note that this was already understood much before the Weihs et al papers. Indeed, Clauser, Horne, and Shimony in \cite{bell_exchange_1985} already noticed that the backward light cones of the events corresponding to the change of measurement settings on the two sides of a Bell experiment will eventually overlap and that if these changes are not driven by a truly random process, they could be due to some factors in this overlap. The factors responsible for the measurement setting on one side could then affect causally the other side. But, they continue
\begin{quote}
We cannot deny such a possibility. But we feel that it is wrong on methodological grounds to worry seriously about it if no specific causal linkage is proposed. In any scientific experiment in which two or more random variables are supposed to be randomly selected, one can always conjecture that some factor in the overlap of the backward light cones has controlled the presumably random choices. But, we maintain, scepticism of this sort will essentially dismiss all results of all scientific experimentation. Unless we proceed under the assumptions	that hidden conspiracies of this sort do not occur, we have abandoned in advance the whole enterprise of discovering the laws of nature by experimentation.
\end{quote}

\section{Superlocality: a bad rationale for QRNGs}\label{sec:qrng}
Despite the above flaws, superlocality is often used to defend the idea that the measurement selection in a Bell experiment should be made using a quantum process. However, even if one adopts blindly superlocality, it provides only a very weak motivation for QRNGs.

It is true that according to our own present understanding a QRNG represents an intrinsically random process and thus can be viewed as a magic box that enables defining unambiguously the moment at which a measurement choice is made. But this is not the right way to look at things. The aim of a Bell experiment is not to confirm our own beliefs about quantum theory; the aim is not to convince a quantum physicist that quantum physics is correct. The aim instead is to rule out a local hidden-variable view. But to arrive at such a conclusion, one should not introduce from the beginning assumptions that may be explicitly refuted by a local hidden-variable model. That the output of a QRNG is intrinsically random is such an assumption. According to a local hidden-variable model, the output of a QRNG is most probably, similarly to the result of any quantum measurement, completely determined by hidden-variables. So, if we plainly adopt the definition of superlocality and use it to discredit experiment involving measurement choices coming from pre-existing lists, pseudo-random generators, coin tosses or Twitter-RNGs, then we should also by the exact same argument discredit experiments using QRNGs. There is no basis to favor one mechanism over another one; they are all on the same footing according to superlocality.

People may say, rightly, that there is still a difference between a QRNG and a classical pseudo-random mechanism. It is true that we do not know for sure whether a QRNG represents a truly random process, but we at least do know for sure that any classical pseudo-random mechanism does admit a completely deterministic explanation and thus cannot be used to close the locality loophole according to superlocality. In the case of a QRNG, we have the benefit of the doubt. It may be that the output of a QRNG is completely deterministic, in which case, the locality loophole is open and nothing can be concluded from the experiment -- as would have happened with a classical pseudo-random choice. But it could also be that the output of a QRNG is indeed truly random, in which case we have ruled out any possible local hidden-variable explanation of our experiment -- something that would not have been possible with a classical pseudo-random choice. In other words, QRNGs can be used to rule out a certain class of local hidden-variable theories, those that posit that the outputs of the QRNGs used in the experiment are really random; other local hidden-variable models, those for which the outputs of the QRNGs are deterministic can simply not be ruled out.\footnote{This point is explicitly made, e.g., in \cite{scheidl_violation_2010} where it is stated that ``the locality and freedom-of-choice loopholes can be closed only within nondeterminism, i.e., in the context of stochastic local realism'' and in \cite{hensen_experimental_2015} where the authors say that their `` observation of a loophole-free Bell inequality violation thus rules out all local realist theories that accept that the number generators timely produce a free random bit''.}

This conclusion is, I think, correct and inescapable if one adopts the definition of superlocality. The problem is that it looks as arbitrary to assume that the output of a QRNG is random according to a local hidden-variable model, than it is to assume that the random boxes in a Twitter-based experiment do not subluminally influence the entangled particles. A local hidden-variable model may (or may not) involve arbitrary hidden subluminal signals between different parts of the experimental set-up and it may (or may not) provide a fully deterministic explanation for the randomness of quantum measurements. The superlocality view is completely asymmetric in that it proceeds from the premise that one should rule out every local hidden-variable model exploiting the first possibility, but is happy not ruling out models exploiting the second possibility. 

Not only there is no good justification for this asymmetric treatment, but if any asymmetry should be introduced, it would be more natural in the other direction! This is because part of the motivation for considering (local or non-local) hidden-variable theories in the first place is precisely to explain in deterministic terms the random nature of quantum theory. Moreover, stochastic local hidden-variable models, in which a specification of the hidden-variables only determines the probability to obtain a certain measurement output, but not a definite one, are more limited than deterministic local hidden-variable models. For instance, even before attempting to model measurements in different directions on two particles in a singlet state, simply reproducing the perfect anti-correlations that arise when performing measurements in the same direction already forces one to consider deterministic models. Many hidden-variable models that have been introduced are indeed deterministic models (famous examples of such models are those of Werner \cite{werner_quantum_1989} and Toner-Bacon \cite{toner_communication_2003}). So stating as a pure matter of principle, as, e.g., in \cite{scheidl_violation_2010}, that deterministic models cannot be tested does not seem a pertinent approach. Furthermore as we have seen in the previous Sections, there are many general families of deterministic local hidden-variable models that can be ruled out through Bell experiments, e.g. those for which the quantum particles are not influenced by or correlated to Twitter-RNGs. 

\section{Pseudo-randomness vs quantum randomness}
So far, I have argued in Sections \ref{sec:plausible} and \ref{sec:conclusive} that a convincing Bell experiment addressing the locality loophole can be done without QRNGs and challenged in Sections \ref{sec:superloc} and \ref{sec:qrng} the superlocality view, according to which QRNGs should necessarily be used. However, even if the motivations for using QRNGs in Bell experiments are misguided, this does not necessarily imply that such experiments are problematic.

Indeed, in the same way that we may find reasonable to assume that the observed quantum systems are not influenced in any relevant way by the content of the random boxes in a Twitter-based experiment, it may be perfectly reasonable to assume that they are not influenced by which path some photon inside a QRNG, out of the zillions of photons present in the lab, takes at the exit of a beam-splitter (even if this decision is completely determined by hidden-variables). This is even more true if as in the experiment of \cite{hensen_experimental_2015} the measurement choice is based on a parity calculation involving 32 successive raw output bits of the QRNG.

The point, however, is that if such an experiment represents a convincing Bell test this is \emph{not because} of the quantum nature of the process involved in selecting the measurement settings. As we have already seen, for most local hidden-variable models of interest, the decision for a photon to take the transmitted or reflected path (or the output of any other quantum measurement) is conceptually no different than for a coin to land heads or tails\footnote{In particular, estimates on the predictability error, on the``freshness" time, or any metrological assurance on the randomness of the QRNGs, as reported in \cite{abellan_generation_2015}, heavily rely on a  quantum modelling of the devices, but may not hold in the context of general local hidden-variable models, and are thus not pertinent to the conclusions that can be drawn from the experiment.	}. But in both case these arbitrary events are unlikely to have any relevant influence on the observed entangled particles. 
 
But then wouldn't it be easier from the implementation point of view to use pseudo-random measurement choices rather than quantum ones, as has been done so far? Actually, this may not only simplify implementations, but also provide a conceptual advantage.

Indeed, when interpreting a Bell experiment, we should weight the possibility  that the random boxes (whether they are Twitter-RNGs or QRNGs) could be correlated to the entangled systems that we are testing. This is similar to an issue that can arise in drug tests. Suppose that we observe people suffering from depression and that we observe two kinds of people: those that take an antidepressant and which have a high rate of recovery and those that don't take any drug and which have a low rate of recovery. We would like to conclude that antidepressants are effective in treating depression. But we do not know what cause people to take a drug or not. It could happen that people which have a tendency to recover naturally from depression also have a higher tendency to take antidepressants. Maybe because they are more resilient, believe deep down that their problem can be cured, that the antidepressants will have some positive effect, etc. And these character treats also explain why they recover quicker from depression. So we are not sure that the data that we observe really establish a causal link between taking antidepressant and recovering from depression. There could be hidden factors (personality treats) that influence the two things and which explains the observed correlations.

So what should we do? We shouldn't leave the decision of taking or not an antidepressant to parameters over which we have no control or are ignorant about, but we should base it on external parameters that we believe have no influence at all on depression. This means that we perform a randomized trial in which we toss coins and decide on this basis who takes an antidepressant and who doesn't. We are then pretty sure that whether someone did or didn't took a drug is uncorrelated to anything about their psychology, health, and so on. This is because we know now exactly what causally determine whether someone takes a drug and this is completely explained by the coin tosses. 

The situation is similar with Bell experiments. In such experiments, we observe a certain type of correlations between the measurement settings and the measurement outputs. But if we use QRNGs to select the measurement settings, we do not really know why one particular setting was chosen and not another one. The QRNGs should be considered as black boxes, potentially described by the same unknown local hidden-variable theories that we are attempting to rule out. It could be in particular that there are hidden factors that explain both the outputs of the QRNGs and the outcomes of the measured entangled particles.

Here is an example of how this could arise. Suppose that there exists a yet undiscovered scalar field in nature, which I call the randomic field, and which takes some value at each space-time point. When we measure a quantum particle, the particle determines the measurement outcome by probing the randomic field. If the local value of the field is 0, the particle outputs something, if it is 1, the particle outputs something else. Prior to a Bell experiment, the underlying randomic field may exhibit long-range correlations that have been established through purely local effects. For instance, a fluctuation of the randomic field at a given point may have propagated as a subliminal wave through space-time and as a result the value of the randomic field at distant space-time points are now correlated. And because the photons in the QRNGs and the entangled particles subject to the Bell test determine their outputs by probing the local value of the randomic field, such outputs can become correlated, even if the QRNGs and entangled photons were initially completely independent. Maybe this explains the Bell violations that we observe in Nature.

Of course, I do not pretend that this is actually a serious explanation of the violation of Bell inequalities. There are probably many good physical reasons why such an explanation cannot work. But I can at least imagine a professional physicist thinking along those lines. The point that I want to make is that since we do not know what causes the output of a QRNG to take a certain value or another, we don't have any good criterion to decide if it is physically plausible or implausible that the observed entangled particles and the QRNGs are correlated.

We should therefore do something analogous to the drug test. We should not base the measurement selection on a black-box that outputs number for which we have no explanation or no control. But we should instead use a mechanism that we completely understand and which as less as possible involves anything deeply related to quantum physics.  Taking a hash function of the latest $10^6$ tweets with the hashtag $\#$quantum is such an example. We then have an explanation for why in the 137th measurement run the measurement settings of Alice was 0: we can deterministically trace this back to the content of the $10^6$ tweets. And it then becomes possible to assess the physical plausibility of some prior correlations between the source of entangled particles and the choice of measurement settings. It is true that some hidden factors could both be responsible for why Graeme Smith tweeted on September 12 ``Snowden leak: NSA would classify 3 logical qubits, 21 physical, plans for 51. $\#$quantum $\#$openscience  @aclu @NSAGov http://bit.ly/1OliJQv'' and why in a Bell test performed a little bit later the entangled particle on Bob's side outputted +1 when measured in the $z$ direction in the 137th measurement run. But I do not imagine any professional physicist entertaining seriously this possibility. 

Let me put this in a more dramatic way. Suppose that it turned out that the Bell violations that we observe in the lab when performing a Bell experiment with QRNGs were only apparent and due to prior correlations between the photons used in the QRNGs and the entangled particles produced by the source. Suppose that this was the explanation for the observed violations. This would of course be a revolutionary discovery and newspapers all around the world would run headlines such as ``Physicists find that quantum particles are correlated to other quantum particles in a mysterious way!''. Now suppose instead that when we do a Bell experiment using Twitter-RNGs, the Bell violations that we observe were due to prior correlations between the content of the tweets and the entangled particles. Suppose that this was the explanation for the observed violations. The headlines would then be ``Physicists discover that Twitter is central to understand quantum effects!'' Which of these two scenarios would you find more surprising?

\section{Random choices on Canary Islands ...}
To finish, let me say a few words about two proposals that aim to provide stronger evidence against local hidden-variable models than tests \`a la Weihs et al. The first one is the experiment by Scheidl et al \cite{scheidl_violation_2010}. In this paper, the authors are worried about the possibility that some factor in the overlap of the backward light cones of the measurement settings could influence both these settings and the entangled particles, a worry that we have already discussed above. The way that this issue is addressed in \cite{scheidl_violation_2010} follows the superlocality paradigm.

The authors thus first claim that if the measurements are selected through some deterministic process, then one cannot close the locality nor the measurement-independence loophole and thus that no conclusive Bell test can be made. They thus rely on QRNGs to choose the measurements and assume in addition that the QRNGs outputs are not deterministically determined by hidden-variables. But if the QRNGs were completely random, everything would be fine. They thus consider an intermediate situation were given a hidden-variable $\lambda$ specifying the output bit $x$ of a QRNG, the output is neither deterministic, $P(x|\lambda)<1 $, nor completely random, $P(x|\lambda)\neq 1/2$. 

They then consider the possibility that some factors in the common past of the QRNGs, could correlate the QRNGs hidden-variables to those characterizing the source of entangled particles, and that -- though these hidden variable do not determine the QRNGs outputs, but only their probabilities -- this could explain the observed Bell violation. For some reason for which no clear justification is given, they further consider that such correlations can only originate from the moment at which the entangled photons are emitted by the source. They thus claim that if things are arranged so that this event is outside the overlap of the backward light cone of the measurement settings, which is achieved in \cite{scheidl_violation_2010} by performing an experiment between two Canary Islands, then the hidden-variables of the QRNGs and of the entangled  particles are not correlated and the ``freedom-of-choice" loophole is closed. 

The problem with the above reasoning and series of assumptions is that they look as arbitrary, if no more, than the ones involved in a Twitter-based experiment. Why is it that the source of entangled particle can only start influencing the QRNGs at the moment at which the entangled pair are emitted by the source but not before? Can't what we call ``photons" according to our present quantum understanding be preceded by scout hidden particles in some other theory? Why if the hidden-variables specify in a completely deterministic way the outcomes of quantum measurements, the authors assume the existence of arbitrary correlations between these hidden-variables, but not if these hidden-variable only partly determine the outcomes? What about models, as the one involving the randomic field discussed above, where hidden-variables determining the ``random" decisions of quantum measurements are not attached to quantum particles but form a background field? They are not excluded by experiments like \cite{scheidl_violation_2010}, yet they seem natural and  leave the ``freedom-of-choice" loophole open.

\section{... or from far-away quasars}
An other idea that has often been put forward to counter possible correlations between the measurement choices and the entangled particles is to use cosmological signals originating from very distant regions of space to choose the measurement settings. This idea has been discussed in great detail in \cite{gallicchio_testing_2014} where the authors propose to use photons originating from quasars on opposite sides of the sky. While experiments like \cite{weihs_violation_1998} and \cite{scheidl_violation_2010} based on QRNGs can be explained by local hidden-variable models if prior correlations between the QRNG outputs and the hidden-variables have been established merely milliseconds before each detector's measurement, the authors of \cite{gallicchio_testing_2014} claim that
their proposal would require that such correlations have been established billions of years earlier, thus involving some conspiracy of cosmic scale. 

This would indeed be true for an experiment with RNGs located billions of light-years from each other. However, in the proposal of \cite{gallicchio_testing_2014}, as in any similar proposal one can think of, the ``random number generators" are not located on distant quasars. Quasar photons must be turned into a bitstream and thus some of their property must be measured. The authors of \cite{gallicchio_testing_2014} propose for instance to measure their arrival time and change the measurement setting based on whether the photons arrive on an even or odd microsecond. A quasar-RNG thus corresponds to a device that extends in space from the quasar photon source to the apparatus used to detect them. But then the two RNGs on the two-side of the Bell experiments are not located billions of light-years away from each other, but only as far away as the two final detectors themselves, typically not more than in a standard Bell experiment.

One could answer that the decisions about whether the two photons arrive on an even or odd microsecond were already taken on the distant quasars and thus that it is fair to assume that the two choices were separated by billions of light-years. But this is true only if one assumes that the photon arrival time can be accurately measured and is not subject to fluctuations modulated by hidden-variables in the vicinity of the telescopes and detectors. Different known physical processes, as explicitly stated in \cite{gallicchio_testing_2014}, could affect the arrival time measurements such as interactions with local medium in the atmosphere or telescopes, or detection of photons of local origin like airglow or light pollution. All these processes could introduce unwanted correlations. More generally, the measurement of the arrival time of quasar photons, or simply whether an incoming photon is detected or not, could also be influenced through yet unknown processes involving  hidden-variables in the vicinity of the detectors. For instance, a photon could decide if he is detected or not by probing a local underlying randomic field or based on whether he arrives on an odd or even microsecond. Such kind of worries may seem paranoid, but they are not that much different than those that lead to the detection loophole, which is taken seriously.

In summary, when one states that in a quasar-based experiment prior correlations could only have been established billions of years earlier, one implicitly relies on several assumptions about the behavior of the telescopes, detectors, etc., and in particular that they do not behave conspiratorially. Such assumptions are perfectly reasonable. But they are essentially no different than those made in a Twitter-RNG experiment, which thus provides the same conclusions at a reduced implementation cost. Note further that the assurance on the randomness of the measurements choices in a quasar-based experiment relies on a heavy and complex modelling involving our present understanding of the physics of photo-detectors, astronomical objects, light refraction in different media, potential noise sources, etc. This probably offers more opportunities for unnoticed hidden influences to enter the picture than in the simpler, and comparatively ``physics-free", Twitter-based experiment.

\section{Final words}
I have argued here, contrarily to claims which are frequently made, that a Bell experiment where the measurement choices are chosen according to some complex pseudo-RNGs is essentially as convincing, or even more, than experiments involving QRNGs \cite{weihs_violation_1998,erven_experimental_2014,hensen_experimental_2015}, QRNGs space-like separated from the entangled particle source \cite{scheidl_violation_2010}, or choices coming from distant cosmic objects \cite{gallicchio_testing_2014}. In each of these cases, some assumptions are needed to rule out local hidden-variable theories and these assumptions are no more natural (sometimes quite to the contrary) than those needed in, say, a Twitter-based experiment. 

Strictly speaking each different experiment rules out a slightly different class of local hidden-variable models. I personally consider that ruling out local hidden-variable models up to some reasonable level of conspiracy is entirely sufficient, which can already be achieved using pseudo-RNG measurement choices. But other people may hold the view that we should attempt to rule out the largest possible class of local hidden-variable models. From this point of view, performing different Bell experiment with distinct ways to generate the measurement settings, or a single experiment where the measurement settings are obtained by combining pseudo, quantum, or cosmic random choices would be justified as it would rule out a class of models strictly larger than any experiment using a single mechanism for selecting the measurements. 

\paragraph{Acknowledgements.}
These notes are based on a talk I gave at the \emph{GISIN'14} workshop in September 2014 and at the \emph{Randomness in Quantum Physics and Beyond} conference in May 2015. I thank Nicolas Gisin for inviting me to the former event and Antonio Ac\'{\i}n to the second one. 

\bibliographystyle{unsrt}
\bibliography{qrng_bell}

\end{document}